# Large-scale 2D Electronics based on Single-layer $MoS_2$ Grown by Chemical Vapor Deposition

H. Wang[1,*,†], L. Yu[1,†], Y.-H. Lee[1], W. Fang[1], A. Hsu[1], P. Herring[1], M. Chin[2], M. Dubey[2], L.-J. Li[3], J. Kong[1], and T. Palacios[1,*]

[1]MIT, 77 Massachusetts Avenue, Cambridge, MA 02139, USA. Tel: +1 (617) 460-7364

[2]United States Army Research Laboratory, 2800 Powder Mill Road, Adelphi, MD 20783-1197, USA.

[3]Institute of Atomic and Molecular Sciences, Academia Sinica, Taipei, 11529, Taiwan

*Corresponding author E-mail: hanw@mtl.mit.edu, tpalacios@mit.edu

## Abstracts

2D nanoelectronics based on single-layer $MoS_2$ offers great advantages for both conventional and ubiquitous applications. This paper discusses the large-scale CVD growth of single-layer $MoS_2$ and fabrication of devices and circuits for the first time. Both digital and analog circuits are fabricated to demonstrate its capability for mixed-signal applications.

## Introduction

A single-molecular layer of $MoS_2$ consists of a layer of Mo atoms sandwiched between two layers of S atoms (Fig. 1(a)). As a 2D material, it shares many properties of the well-known graphene such as excellent mechanical flexibility and high thermal stability up to 1090 °C. While with a direct bandgap of 1.8 eV [1], it also overcomes the key shortcomings of graphene for electronic applications - the lack of clear current saturation and pinch-off [2]. 2D electronics based on $MoS_2$ represents the ultimate limit of scaling in the vertical direction, which is necessary to keep pushing Moore's Law. Its relatively wide bandgap is also advantageous over silicon for suppressing the source-to-drain tunneling at the scaling limit of transistors [3]. Moreover, 2D nanoelectronics based on $MoS_2$ and other transition metal dichalcogenides (TMD) materials are attractive as high-mobility options in the emerging field of large-area low-cost electronics that is currently dominated by low-mobility amorphous silicon and organic semiconductors (<10 $cm^2$/Vs) [2, 4, 5]. Single-layer $MoS_2$ can also complement graphene [6, 7, 8] to build flexible digital and mixed-signal circuits. This paper addresses several key challenges in the development of 2D nanoelectronics on $MoS_2$ and TMD materials in general. The main contributions are three-fold. First, large-area single-layer $MoS_2$ material is grown by chemical vapor deposition (CVD) that makes the wafer-scale fabrication of $MoS_2$ devices and circuits possible for the first time. Second, the top-gated transistors, fabricated for the first time on single-layer $MoS_2$ grown by CVD, show multiple state-of-the-art characteristics, such as record mobility for CVD $MoS_2$ (>40 $cm^2$/V.s), ultra-high on/off current ratio (>$10^8$), record current density (close to 20 μA/μm), current saturation and the first GHz RF performance. Finally, the first fully integrated digital and analog circuits based on $MoS_2$ are constructed to demonstrate its capability for both logic and mixed-signal applications.

† H. W. and L. Y. contributed equally to this work.

## CVD Growth and Device/Circuit Fabrication

Fig. 1(b) shows the experimental set-up for the CVD synthesis of single-layer $MoS_2$, using S and $MoO_3$ as the precursor, and perylene-3,4,9,10-tetracarboxylic acid tetrapotassium salt (PTAS) as the seed [9]. Optical micrograph and Raman spectroscopy confirm that single-layer $MoS_2$ with uniformity greater than 95% can be grown at centimeter-scale, which is only limited by the size of the reaction furnace. The high quality of the material is also evidenced by the outstanding photoluminescence peaks (Fig. 2) derived from its 1.8 eV direct bandgap and the highly ordered lattice structure that can be observed under high resolution TEM (Fig. 3). AFM data confirms the thickness of the material is about 6.9 Å, typical of single-layer $MoS_2$ (Fig. 4). The devices/circuits fabrication starts from the formation of ohmic contacts by depositing a 5nm Ti/ 50 nm Au metal stack using e-beam evaporation. Device isolation is achieved by $O_2$ plasma etching. The top gate dielectric consisting of 30 nm $HfO_2$ is then deposited by ALD at 170°C. To fabricate discrete transistors, the top gate is then formed by depositing 50 nm Pd. For the construction of integrated logic circuits, via holes are created on $HfO_2$ before the formation of the gate metal layer (M2, Fig. 5) using reactive ion etching (RIE) with $BCl_3/Cl_2$ gas chemistry to connect M2 to M1. Fig. 6(a) and shows the chip fabricated using single-layer $MoS_2$ grown on $SiO_2$. Fig. 6(b) shows a top view of the various integrated device and circuit arrays on the chip of Fig. 6(a).

## FETs Based on CVD Grown Single-Layer $MoS_2$

Fig. 8(a) shows the output characteristics of the fabricated single-layer $MoS_2$ FET with $L_G$=$L_{DS}$=1 μm. The device shows clear current saturation, which is due to the pinch-off of the channel at the drain side of the gate at high $V_{DS}$ since the saturation voltage $V_{sat}$ closely matches the gate overdrive $V_{GS}$-$V_t$. This is the first demonstration of current saturation in top-gated $MoS_2$ FETs, a crucial device characteristic that has been missing in previous reports of $MoS_2$ FETs [10]. The maximum on-state current reaches 16 μA/μm at $V_{DS}$=5 V and $V_{TG}$=2 V. The threshold voltage is at -2 V (Fig. 8(b)), indicating that the material is unintentionally doped n-type during the growth and fabrication process. The peak transconductance is around 3 μS/μm. The classical drift-diffusion model [11] accurately fits the device characteristics (Fig. 8(a,b)). Due to a larger bandgap than Si and excellent electrostatic control of 2D electronics, the device exhibits a remarkable on/off current ratio exceeding $10^8$, giving the material great potential for ultra-low power applications such as driving circuits for flat panel display, where most of the incumbent materials - organics and amorphous Si - have mobility below 1 $cm^2$/V.s due to their intrinsic disorder and tunneling-based transport mechanism. In contrast, carrier mobility in polycrystalline CVD single-layer $MoS_2$ is extracted to be above 40 $cm^2$/V.s at 300 K (Fig. 7) while single-crystalline exfoliated multi-layer $MoS_2$ shows mobility around 150 $cm^2$/V.s at 300 K. Such mobility allows $MoS_2$ FETs to operate even at GHz frequency. Fig. 8(c) shows the first RF performance characterization of $MoS_2$ FETs. This device with $L_G$=300 nm shows an $f_T$ of 900 MHz and $f_{max}$ of 1 GHz, giving $MoS_2$ the potential to enable high performance RF circuits on bendable substrates, such as flexible RFID tags.

### Integrated Logic Circuits based on Single-Layer $MoS_2$

A key contribution of this paper is the demonstration of large-scale fabrication of integrated circuits based on CVD single-layer $MoS_2$. Fig. 9 and 10 show fully integrated inverters and NAND gates in depletion mode configuration. The inverters can operate under a wide range of $V_{dd}$ from 0.5 V to 5V with peak voltage gain close to 20 at $V_{dd}$=5 V (Fig. 9(c)). Fig. 10(c) shows the NAND gate performing logic function on two input signals. This demonstration of an NAND gate, one of the two types of universal logic gates (the other being NOR), shows that it is possible to realize any Boolean functions on CVD $MoS_2$ thin film.

### Integrated Mix-Signal Circuits based on Single-Layer $MoS_2$

Finally, to demonstrate the mixed-signal capability of $MoS_2$, a 1-bit analog-to-digital converter (ADC) is constructed (Fig. 11(a)) based on a matched long-tail pair differential amplifier (Fig. 11(b)). Using the differential gain of the long-tail pair, the ADC converts a 1 kHz sinusoidal signal to a square wave, which is essentially a digital output with two logic levels. Any part of the input signal above (or below) the reference voltage $V_{ref}$ is represented by logic level 0 (or 1) in the output signal $V_1$-$V_2$ (Figure 11(c)). The successful operation of the ADC also demonstrates the excellent matching between the characteristics of the various transistors used in the circuit, necessary for a successful differential amplifier.

### Conclusion

Integrated devices and circuits based on large-scale single-layer $MoS_2$ grown by CVD are presented for the first time. The transistors fabricated on this material demonstrate excellent characteristics such as record mobility for CVD $MoS_2$, ultra-high on/off current ratio, record current density and GHz RF performance. The demonstration of both digital and analogue circuits shows the remarkable capability of this single-molecular-layer thick material for mixed-signal applications, offering scalable new materials that can combine silicon-like performance with the mechanical flexibility and integration versatility of organic semiconductors.

### Acknowledgements

This work has been partially supported by the ONR Young Investigator Program and the Army Research Laboratory.

### References

[1] Splendiani, A. *et al.* Emerging Photoluminescence in Monolayer $MoS_2$. *Nano Lett.* **10**, 1271–1275 (2010).

[2] Wang H., et. al. Integrated Circuits Based on Bilayer $MoS_2$ Transistors, *Nano Lett.*, 2012, **12** (9), pp 4674–4680.

[3] Wang, J. & Lundstrom, M. Does source-to-drain tunneling limit the ultimate scaling of MOSFETs? *IEEE Tech. Dig. IEDM* 707 – 710 (2002).

[4] Street, R. A. Thin-Film Transistors. *Advanced Materials* **21** (2009).

[5] Forrest, S. R. The path to ubiquitous and low-cost organic electronic appliances on plastic. *Nature* **428**, 911–918 (2004).

[6] Wang, H., Nezich, D., Kong, J. & Palacios, T. Graphene Frequency Multipliers. *IEEE Electron Device Letters* **30**, 547–549 (2009).

[7] Wang, H., Hsu, A., Wu, J., Kong, J. & Palacios, T. Graphene-Based Ambipolar RF Mixers. *IEEE Electron Device Letters* **31**, 906–908 (2010).

[8] Lin, Y.-M. *et al.* Wafer-Scale Graphene Integrated Circuit. *Science* **332**, 1294 –1297 (2011).

[9] Lee, Y.-H., *et. al.* Synthesis of Large-Area $MoS_2$ Atomic Layers with Chemical Vapor Deposition. *Advanced Materials* **24**, 17, 2320–2325 (2012).

[10] Radisavljevic, B., Radenovic, A., Brivio, J., Giacometti, V. & Kis, A. Single-layer $MoS_2$ transistors. *Nature Nanotechnology* **6**, 147–150 (2011).

[11] Tsividis, Y. & McAndrew, C. *Operation and modeling of the MOS transistor*. (Oxford University Press, USA: 2010).

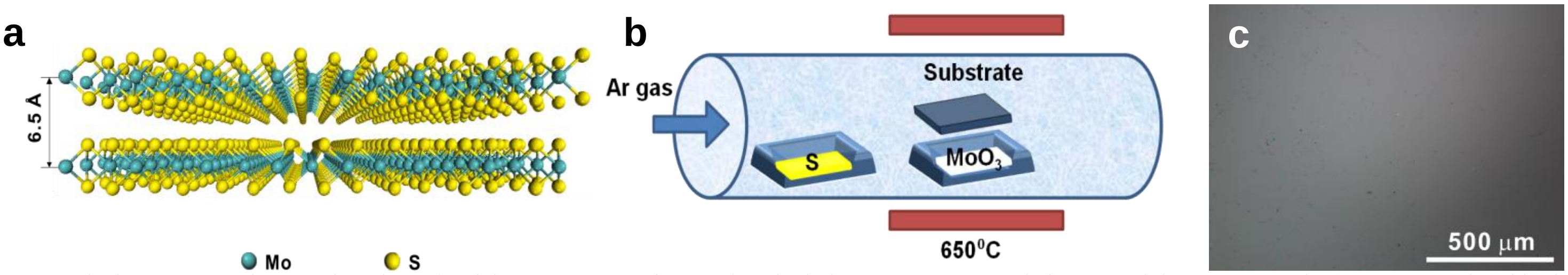

**Fig.1** (a) lattice structure of $MoS_2$. (b) Schematic of the CVD process for growing single-layer $MoS_2$. A ceramic boat containing $MoO_3$ powder (0.02 g) was placed in the center of the quartz tube and a second ceramic boat with sulfur powder (0.01g) was placed upstream. The substrate was pretreated with a droplet of aqueous PTAS solution and then mounted on top of the boat containing $MoO_3$ powder. During the growth, the furnace was heated to $650^0$C in Argon environment. The sulfur vapor was carried by the Argon gas and reduced the $MoO_3$ powder to form volatile suboxide $MoO_3$-x, which diffused to the substrate and was further reduced by sulfur vapor to form $MoS_2$ films. The same approach can also be used to grow other TMD semiconductors such as $WS_2$, $MoSe_2$ and $WSe_2$. (c) Optical micrographs of single-layer $MoS_2$ sheets grown by this method. $MoS_2$ obtained shows great uniformity with more than 95% of the area covered by single-layer $MoS_2$. This is the first time high quality single-layer $MoS_2$ can be grown at wafer scale. Since the material can be grown directly on any insulating substrates that are stable at the growth temperature such as $SiO_2$ and sapphire, no transfer step is required. The material is hence free of wrinkles, which are common sources of mobility degradation often associated with the transfer of CVD grown graphene.

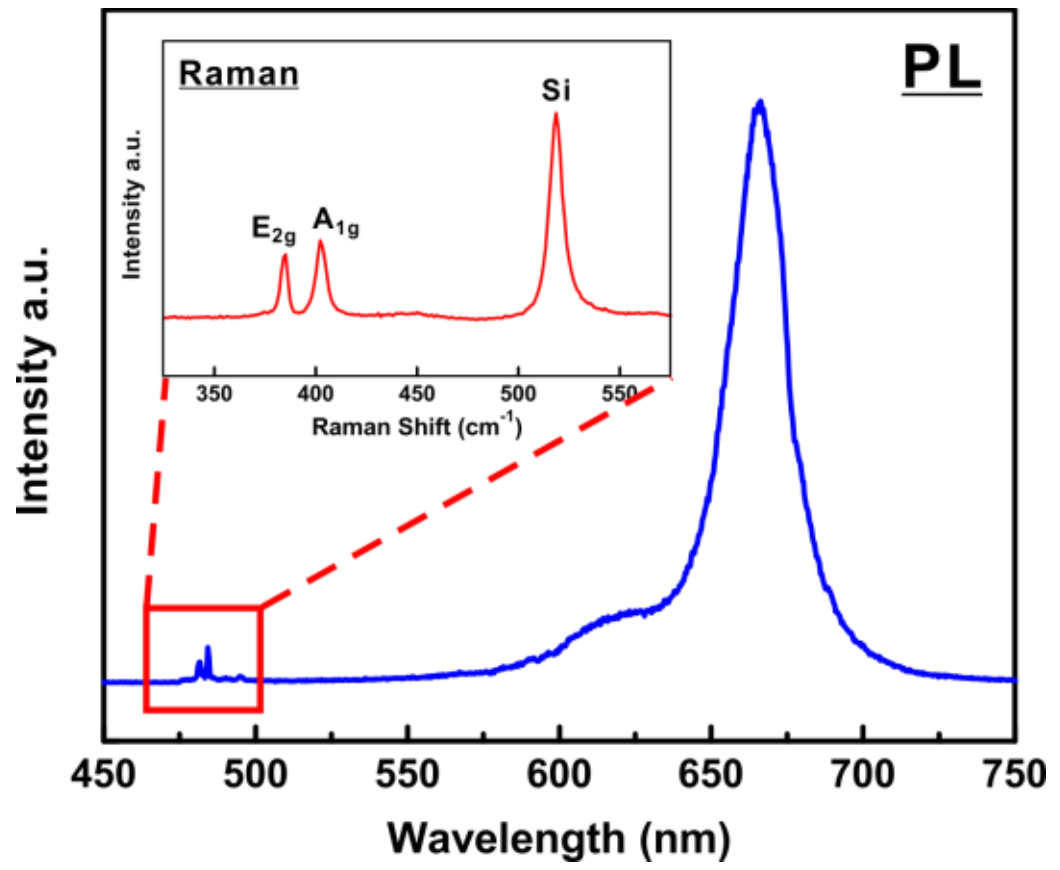

**Fig.2** Photoluminescence (PL) of the single-layer $MoS_2$ grown by CVD. Inset: the corresponding Raman spectra. Both Raman and PL experiments were performed in a confocal spectrometer using a 473 nm excitation laser.

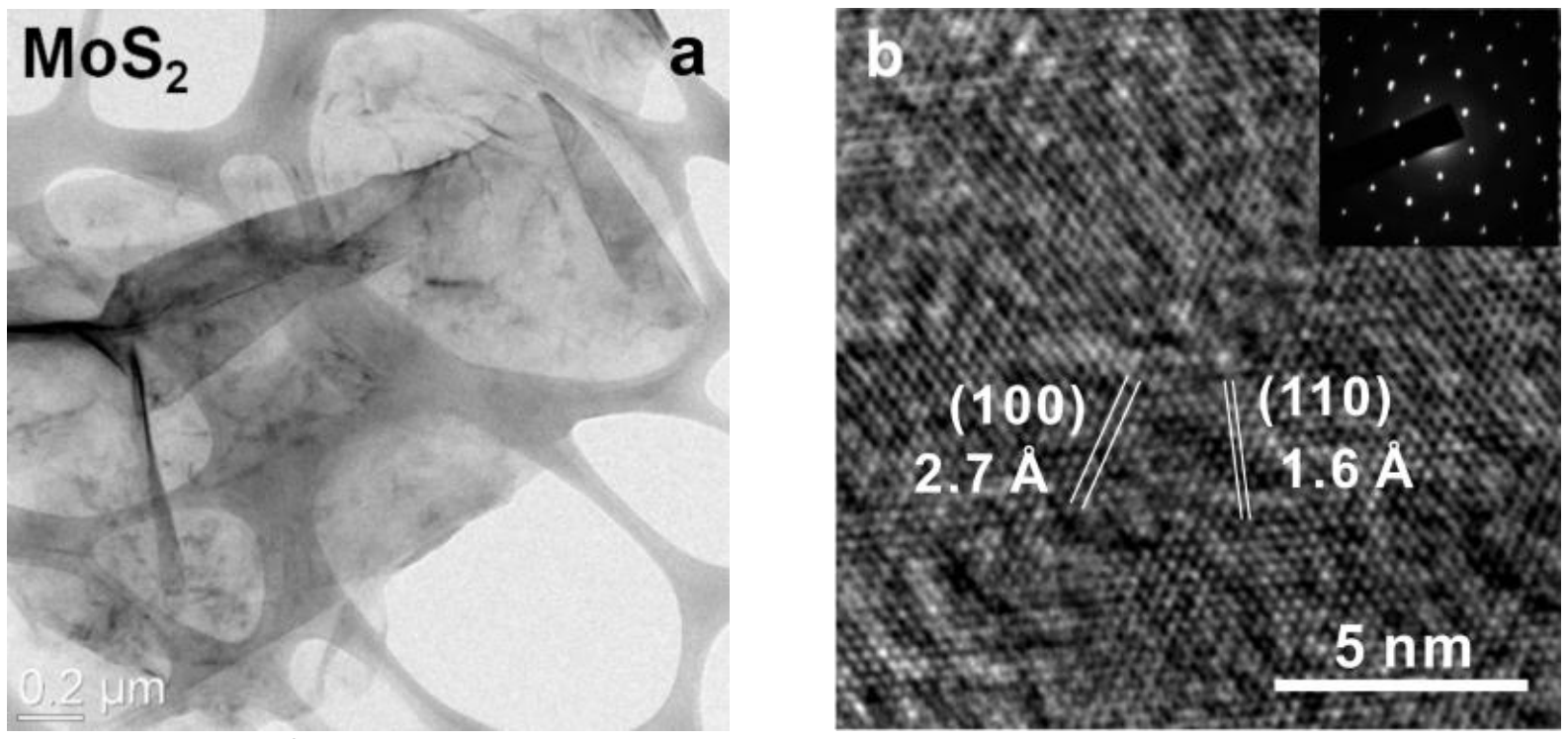

**Fig.3** (a) TEM image of single-layer $MoS_2$ monolayer grown by CVD. (b) HRTEM image and the corresponding selected area electron diffraction (SAED) pattern (Inset) show the hexagonal crystal nature and the lattice spacing of 0.27 and 0.16 nm assigned to the (100) and (110) planes of $MoS_2$.

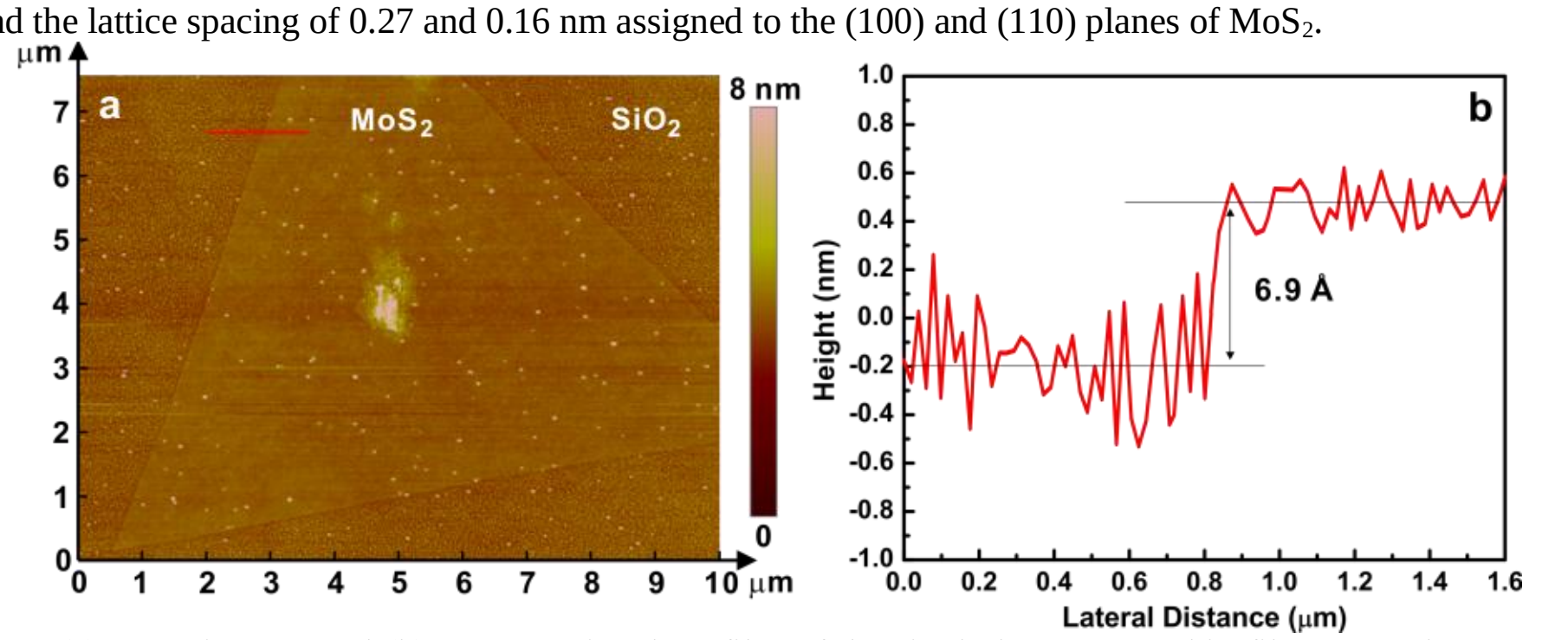

**Fig.4** (a) AFM images and (b) cross-sectional profiles of the single-layer $MoS_2$ thin film grown by CVD. The coated PTAS seeds act as a nucleation site to initiate the growth. In this sample, the CVD growth process is intentionally stopped before the growth initiated from neighboring seeds merges. The AFM image hence shows the single-layer $MoS_2$ grown in an equilateral-triangle pattern that is determined by its underlying lattice structure. The CVD single-layer $MoS_2$ has a thickness of 6.9 Å, typical for single-layer $MoS_2$ sheets.

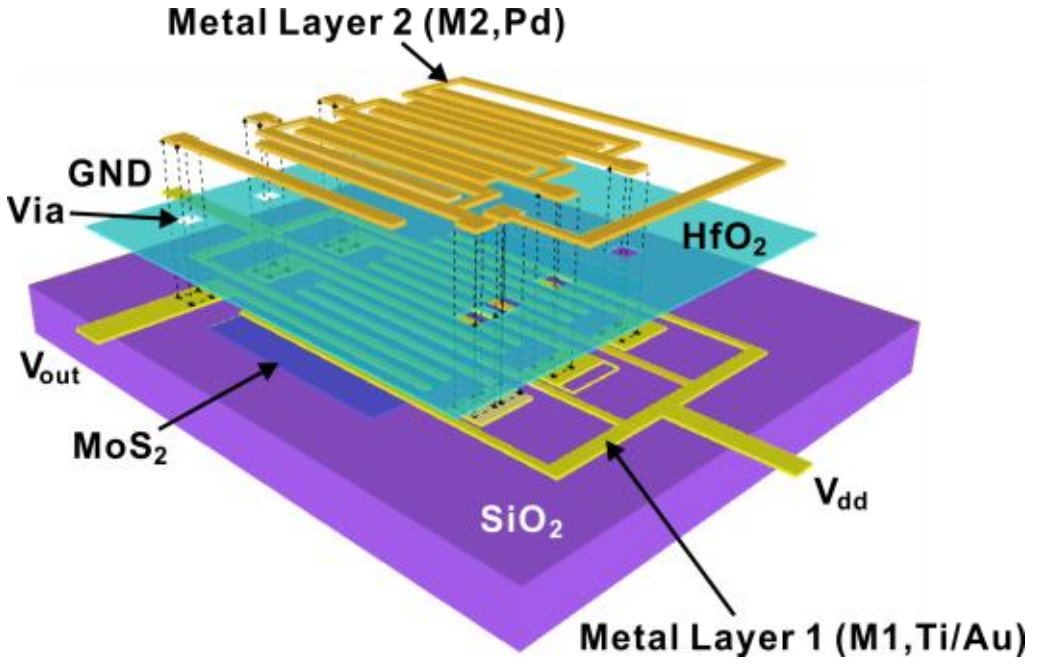

**Fig. 5** Schematic illustration of the fabrication process for building transistors and integrated circuits on $MoS_2$.

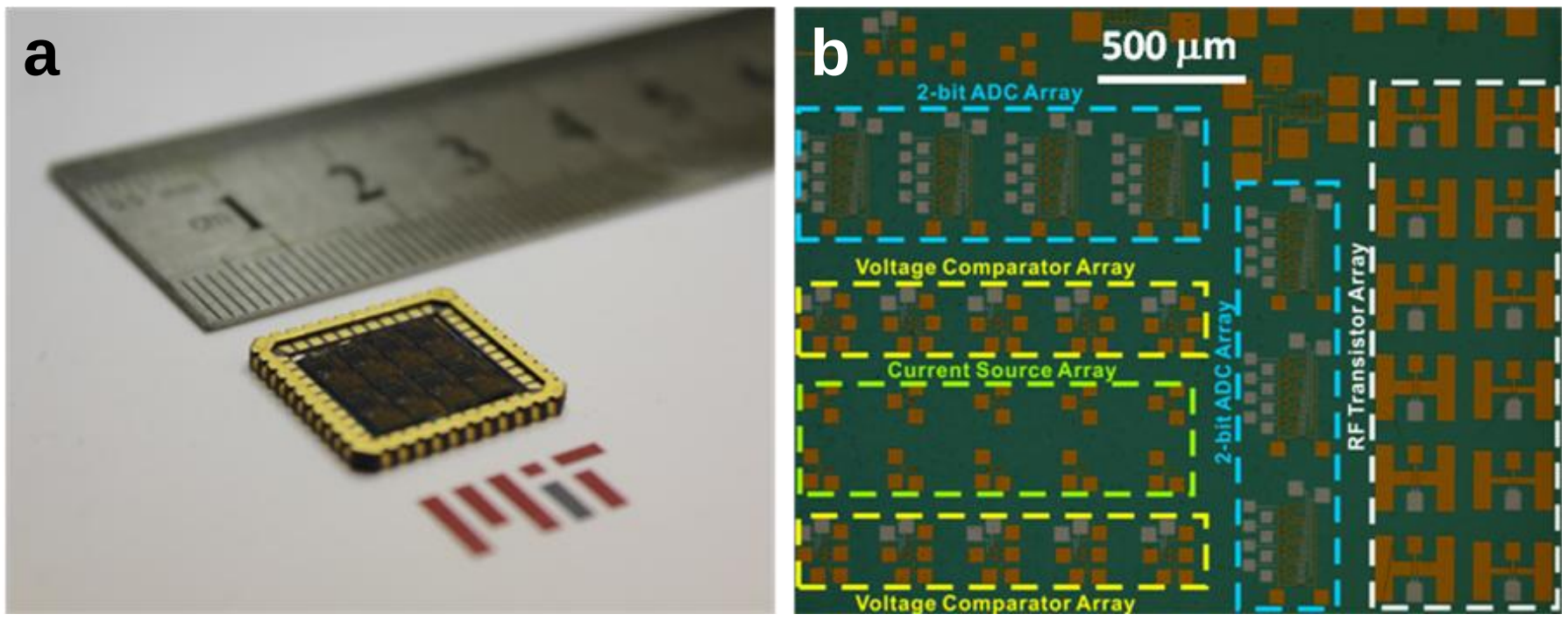

**Fig.6** (a) Transistors and integrated circuits fabricated at the wafer scale for the first time using single-layer $MoS_2$ grown directly on $SiO_2$/Si substrate by CVD methods. (b) Optical micrograph of the chip fabricated using CVD single-layer $MoS_2$ grown on $SiO_2$/Si substrate, showing arrays of RF transistors, currents sources, 1-bit and 2-bit analog-to-digital converters (ADC).

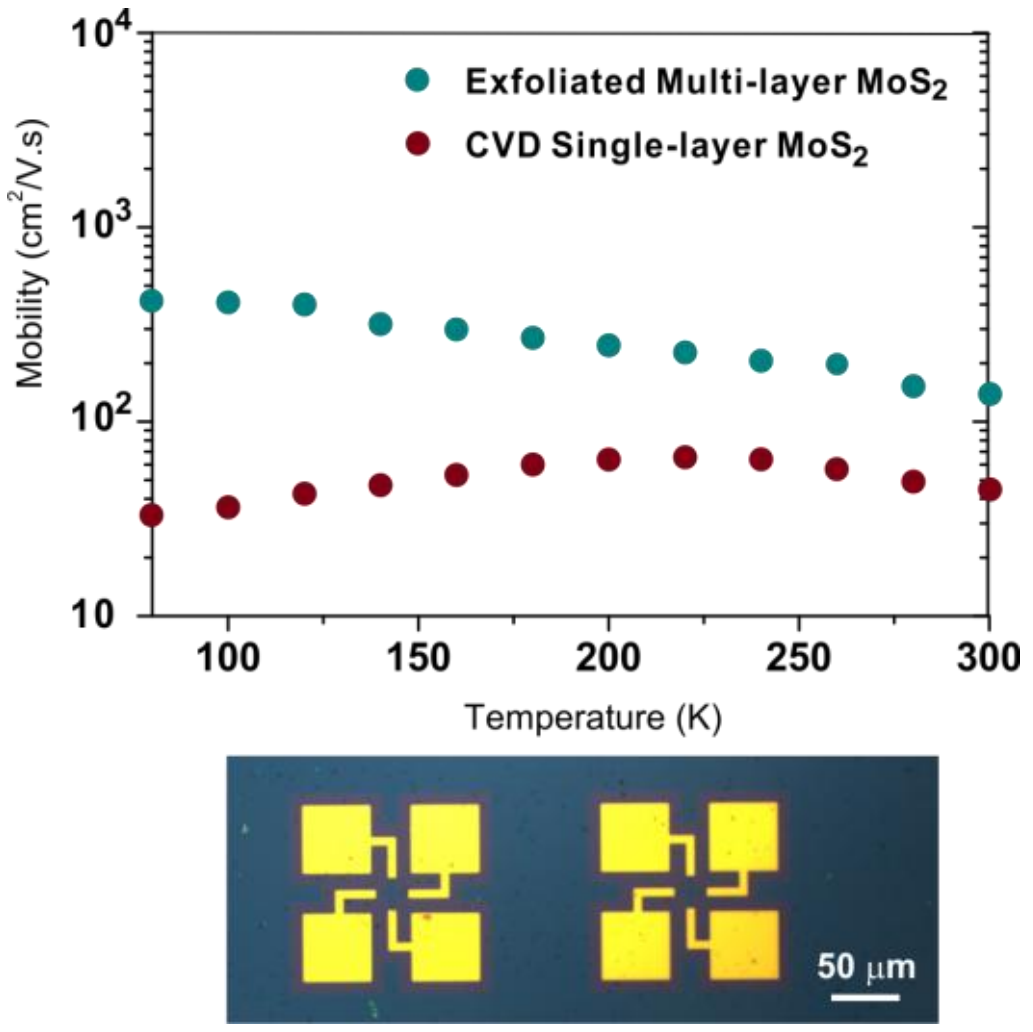

**Fig.7** Hall mobility of exfoliated multi-layer $MoS_2$ and CVD single-layer $MoS_2$ and their temperature dependence. At high temperature, the mobility in both samples are limited by phonon scattering. At lower temperatures, the exfoliated multi-layer sample shows improved mobility due to reduced phonon scattering while the CVD singl-layer sample shows reduced mobility below 200 K, which can be explained by the fact that in this polycrystalline CVD single-layer sample, the transport at low temperature is dominated by charge carrier hopping through localized states.

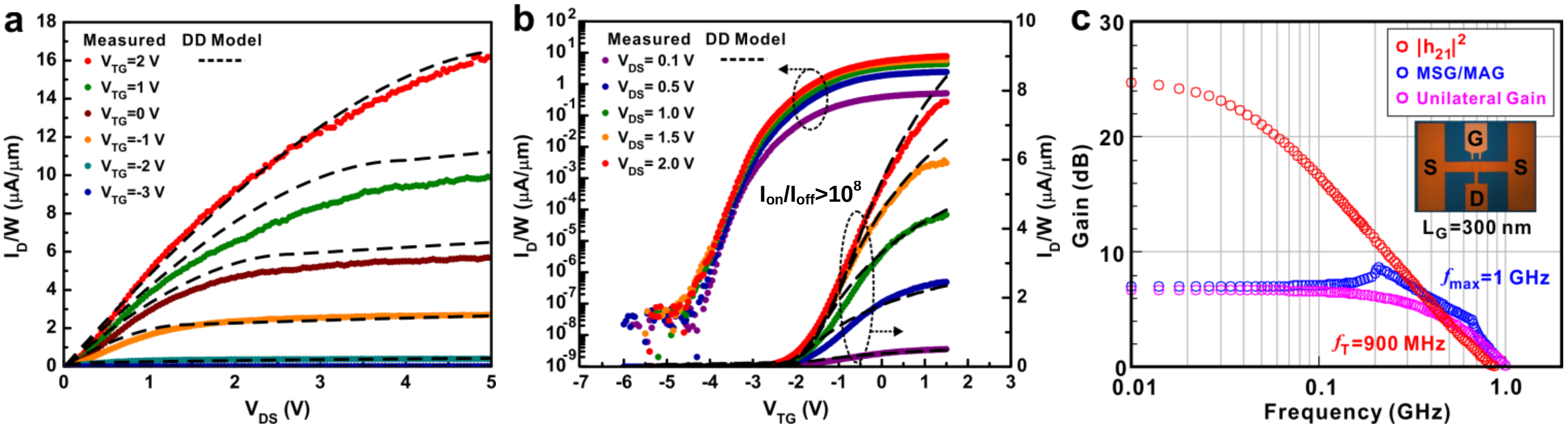

**Fig.8** (a) Output characteristics of the FET fabricated on CVD single-layer $MoS_2$. The device shows excellent current saturation. $L_G$=$L_{DS}$=1 μm. $V_{BG}$=0 V. (b) Transfer characteristics in linear scale (right y-axis) and log scale (left y-axis). The on/off current ratio of the device exceeds $10^8$, making these devices ideal for ultra-low power applications. The subthreshold swing is 110 mV/dec. $V_{BG}$=0 V. The classical drift-diffusion (DD) model (black dotted lines) [7] gives an excellent fit to the data in (a) and (b), indicating that the carrier transport in these devices are dominated by drift-diffusion and current saturation in (a) is due to channel pinch-off at the drain side of the gate, i.e. $V_{sat}$=$V_{TG}$-$V_t$, all similar to conventional long channel MOSFETs. (c) First RF characteristization of a $MoS_2$ FET. The measured device has $L_G$=300 nm. $f_T$=900 MHz and $f_{max}$=1 GHz. This demonstrates the high frequency potential of $MoS_2$ 2D-electronics compared to organic semiconductors and amorphous Si, opening doors for applications such as high performance RFID tags on flexible substrates. It should be noted that this measurement is performed in high vacuum (~$10^{-6}$ Torr) to reduce hysteresis.

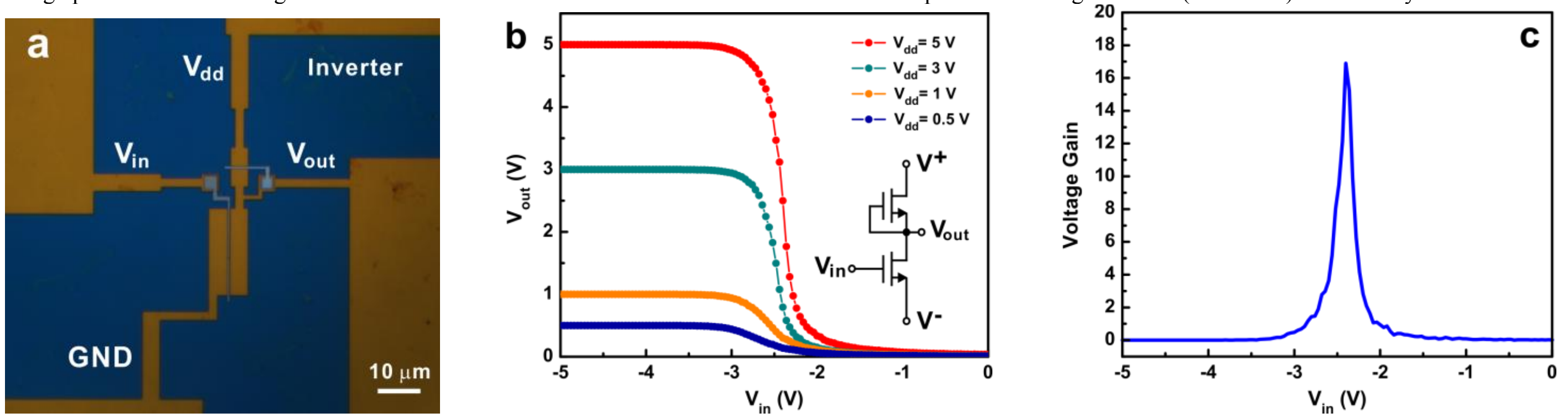

**Fig. 9** (a) Optical micrograph of a fully integrated inverter constructed on single-layer $MoS_2$ (b) Input-output characteristics of the inverter as a function of supply voltage $V_{dd}$. The inverter operates on a wide range of $V_{dd}$ from 0.5 V to 5V. (c) Voltage gain of the inverter. At $V_{dd}$=5 V, the peak voltage gain of the inverter is close to 20.

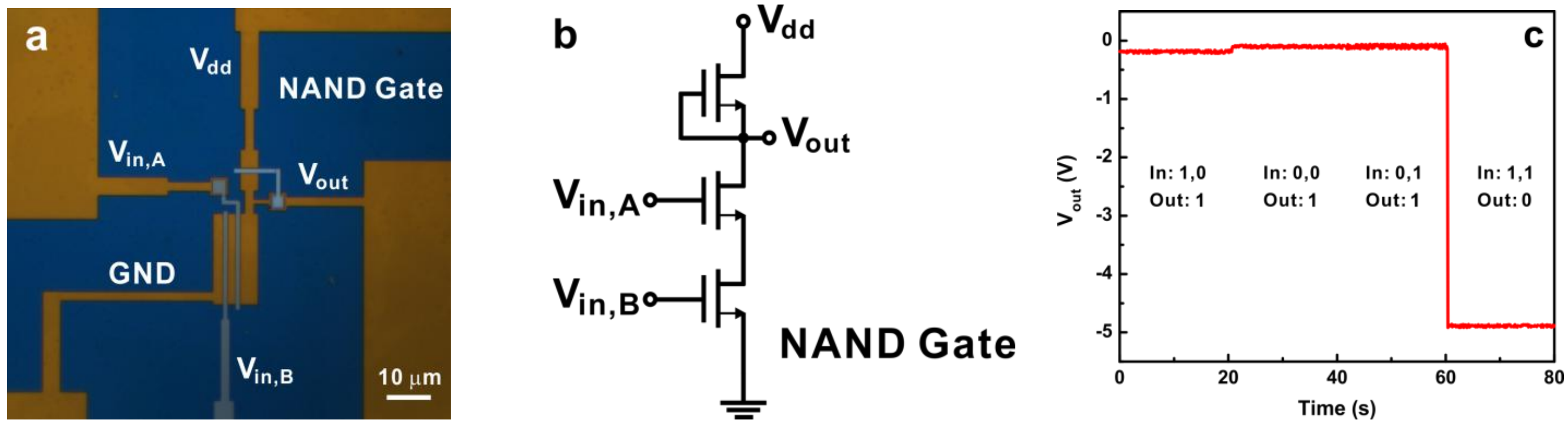

**Fig.10** (a) Optical micrograph of a fully integrated NAND gate constructed on single-layer $MoS_2$. (b) Schematic circuit for the NAND gate. (c) Output voltage of the NAND gate for four different input states: (1,0), (0,0), (0,1), and (1,1). A low voltage of -5 V represents a logic state 0 and a voltage close to 0 V represents logic state 1. This data demonstrates the stable NAND gate operation of this single-layer $MoS_2$ circuit. A NAND gate is one of the two basic logic gates (the other being NOR gate) with universal functionality. Any other type of logic gates (AND, OR, NOR, XOR, etc.) can then be constructed with a combination of NAND gates. Hence, this first demonstration of a NAND gate on CVD $MoS_2$ shows that it is possible to fabricate any kind of digital integrated circuit with $MoS_2$ thin film.

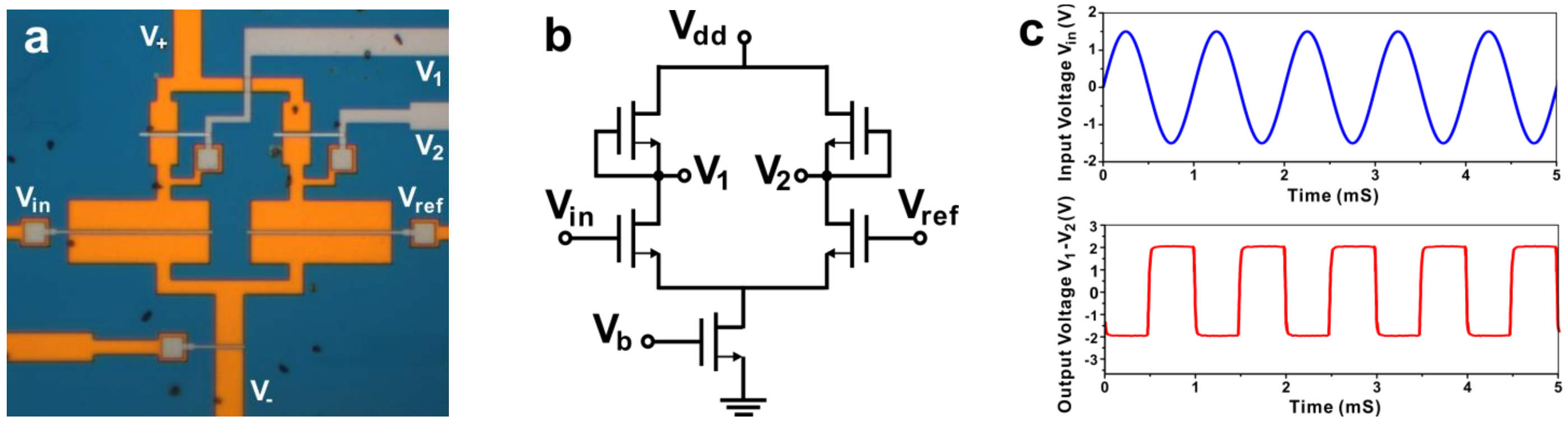

**Fig.11** (a) Optical micrograph of a fully integrated voltage comparator (1-bit ADC) constructed on single-layer $MoS_2$. (b) Schematic circuit for the ADC. (c) AC coupled oscilloscope reading of the input and output signals (1 kHz) of the ADC. The ADC is essentially a voltage comparator based on a differential amplifier with matched long-tail pair configuration, which requires well-matched characteristics in the transistors involved. When the input voltage $V_{in}$ is above (or below) the reference voltage $V_{ref}$, the output voltage is at logic level 0 (or logic level 1). The high differential gain converts the analog input to a digital square wave. This circuit demonstrates for the first time the mixed-signal capability in circuits based on $MoS_2$, or any 2D material in general including graphene. The measurement is done at room temperature under high vacuum (~$10^{-6}$ Torr).